\documentclass[journal,10pt,draftclsnofoot,onecolumn]{IEEEtran}
\ifCLASSINFOpdf
  \usepackage[pdftex]{graphicx}
\else
  \usepackage[dvips]{graphicx}
\fi
\usepackage[caption=false,font=footnotesize]{subfig}
\usepackage{tabularx}
\newcolumntype{Y}{>{\centering\arraybackslash}X}
\usepackage{amssymb}
\usepackage{amsmath}
\usepackage{times}
\usepackage{amssymb}
\usepackage{amsmath}
\usepackage{xspace}
\usepackage{cite}
\usepackage{color}
\usepackage[noend,ruled,linesnumbered]{algorithm2e}
\usepackage{url}

\newtheorem{theorem}{Theorem}

\newtheorem{lemma}[theorem]{Lemma}

\newcommand{\CH}{{\ensuremath{\mathcal{C}}}\xspace}
\newcommand{\NCH}{{\ensuremath{\overline{\mathcal{C}}}}\xspace}
\newcommand{\VRF}{{\ensuremath{\mathcal{V}}}\xspace}
\newcommand{\Wp}{{\ensuremath{\mathit{WP_{\CH}}}}\xspace}
\newcommand{\Wb}{{\ensuremath{\mathit{WB_\mathcal{A}}}}\xspace}
\newcommand{\Wc}{{\ensuremath{\mathit{WC_\mathcal{T}}}}\xspace}
\newcommand{\Mp}{{\ensuremath{\mathit{MP_\mathcal{W}}}}\xspace}
\newcommand{\Vc}{{\ensuremath{\mathit{MC_{\VRF}}}}\xspace}
\newcommand{\Ac}{{\ensuremath{\mathit{MC_\mathcal{A}}}}\xspace}
\newcommand{\Mb}{{\ensuremath{\mathit{MB_\mathcal{R}}}}\xspace}
\newcommand{\gamethrees}{{\ensuremath{0:n}}\xspace}
\newcommand{\modelone}{{\ensuremath{\cal{R}_{\rm m}}}\xspace}

\newcommand{\myboldmath}{}
\newcommand{\defn}[1]{{\textit{\textbf{\myboldmath #1}}}}

\usepackage[normalem]{ulem}
\newcommand{\mig}[1]{\textcolor{blue}{#1}}
\newcommand{\cg}[1]{\textcolor{green}{#1}}
\newcommand{\remove}[1]{}

\begin{document}
%
\title{Multi-round Master-Worker Computing:\\ a Repeated Game Approach~\thanks{
This work has been supported in part by the National Science Foundation (CCF-1114930), Kean University UFRI grant,
Ministerio de Econom\'ia y Competitividad grant TEC2014-55713-R, and Regional Government of Madrid (CM) grant Cloud4BigData (S2013/ICE-2894, cofunded by FSE \& FEDER).
}}

\author{\IEEEauthorblockN{Antonio Fern\'andez Anta}\\
\IEEEauthorblockA{IMDEA Networks Institute,
Madrid, Spain,
antonio.fernandez@imdea.org}\\
\and
\IEEEauthorblockN{Chryssis Georgiou}\\
\IEEEauthorblockA{University of Cyprus,
Nicosia, Cyprus,
chryssis@cs.ucy.ac.cy}\\
\and
\IEEEauthorblockN{Miguel A. Mosteiro, Daniel Pareja}\\
\IEEEauthorblockA{Kean University,
Union, NJ, USA,
\{mmosteir,parejad\}@kean.edu}
}


%


\maketitle

\begin{abstract}
We consider a computing system where a master processor assigns tasks for execution to worker processors through the Internet. We model the workersÕ decision of whether to comply (compute the task) or not (return a bogus result to save the computation cost) as a mixed extension of a strategic game among workers. That is, we assume that workers are rational in a game-theoretic sense, and that they randomize their strategic choice. Workers are assigned multiple tasks in subsequent rounds. We model the system as an infinitely repeated game of the mixed extension of the strategic game. In each round, the master decides stochastically whether to accept the answer of the majority or verify the answers received, at some cost. Incentives and/or penalties are applied to workers accordingly. 

Under the above framework, we study the conditions in which the master can reliably obtain tasks results, exploiting that the repeated games model captures the effect of long-term interaction. That is, workers take into account that their behavior in one computation will have an effect on the behavior of other workers in the future. Indeed, should a worker be found to deviate from some agreed strategic choice, the remaining workers would change their own strategy to penalize the deviator. Hence, being rational, workers do not deviate.  

We identify analytically the parameter conditions to induce a desired worker behavior, and we evaluate experimentally the mechanisms derived from such conditions. We also compare the performance of our mechanisms with a previously known multi-round mechanism based on reinforcement learning.\vspace{.5em}\vspace{.5em}\\
\end{abstract}


%
\IEEEpeerreviewmaketitle

\section{Introduction}
\label{sec:intro}

\noindent{\em Motivation and prior work.}
The processing power of top supercomputers has reached speeds in the order of PetaFLOPs~\cite{top500}. However, the high cost of building and
maintaining such multiprocessor machines makes them accessible only to large companies and institutions. Given the 
drastic increase of the demand for high performance computing, Internet-based computing has
emerged as a cost-effective alternative. One could categorize Internet-based computing into two categories: {\em administrative} computing and
{\em master-worker} computing. 
In the first one, the computing elements are under the control of an administrator. Users of such infrastructure have to bear the cost of access, which depends on the quality of service they require.
Examples of such system
are {\em Grid} and {\em Cloud} computing. Master-working computing (also known as Desktop Grid Computing) exploits the  
growing use of personal computers and their capabilities (i.e. CPU and GPU) and their high-speed access to the Internet, in providing
an even cheaper high performance computing alternative. In particular, personal computing devices all around the world are accessed through the Internet and are used for computations; these devices are called {\em workers} and the {\em tasks} (computation jobs) are assigned 
by a {\em master} entity (the one that needs the outcome of the tasks' computation). 
At present, Internet-based master-worker computing is mostly embraced by the scientific community in the form of volunteer computing, where computing resources are volunteered by the public to help solve scientific problems. Among the most popular volunteering projects is SETI@home~\cite{SETI} running on the BOINC~\cite{boinc} platform. A profit-seeking computation platform has also been developed by Amazon, called  Mechanical Turk~\cite{turk}\footnote{Although in Amazon's Mechanical Turk many tasks are performed by humans, even such cases can be seen as a computational platform, one where processors are indeed humans.}. Although the potential is great, the use of master-worker computing is limited by the untrustworthy nature of the workers, who might report incorrect results~\cite{boinc,Heienetal09,volunteer,Kondoetal2007}. In SETI, the master attempts to minimize
the impact of these incorrect results by assigning the same task to several workers and comparing their outcomes (i.e., redundant task allocation is employed~\cite{boinc}).  

Prior work, building on redundant task allocation, has considered different approaches in increasing the reliability of master-worker computing~\cite{Sarmenta02,ALEX,PPL12,CCS,FGM:mechdesgnConf,ccpe,IEEETC14}. One such approach is to consider workers to be 
{\em rational}~\cite{Halp06,UDC} in a game-theoretic sense, that is, each worker is selfish and decides whether to truthfully compute and return the correct result or return a bogus result, based on the strategy that best serves its self-interest (increases its benefit).
The rationality assumption can conceptually be justified by the work of Shneidman and Parkes~\cite{rational} where they reason on 
the connection of rational players (of Algorithmic Mechanism Design) and workers in realistic P2P systems. 
Several incentive-based algorithmic mechanisms have been devised, e.g.,~\cite{CCS,FGM:mechdesgnConf,IEEETC14,ccpe,plosone}, that employ reward/punish schemes to ``enforce'' rational workers to act correctly, and hence having the master reliably obtain correct task results. Most of these mechanisms are {\em one-shot} in the following sense: in a round, the master sends a task to be computed to a collection of workers, and the mechanism, using auditing and reward/punish schemes guarantees (with high probability) that the master gets the correct task result.
For another task to be computed, the process is repeated 
but without taking advantage of the knowledge gained.

The work in~\cite{ccpe} takes advantage of the repeated interactions between the master and the workers, by studying the {\em dynamics of evolution}~\cite{MSJ82} of such master-worker computations through {\em reinforcement learning}~\cite{RLbook} where both the master and the workers adjust their strategies based on their prior interaction. The objective of the master is to reach 
a state in the computation after which it always obtains the correct results (called eventual correctness), while the workers attempt to 
increase their benefit. 
Roughly speaking, in each round a different task is assigned to the workers, and when the master collects
the responses it decides with some probability to verify these answers or not; verification is costly to the master. 
Each worker decides with some probability to cheat, that is, return a bogus result without computing the task (to save the cost of doing so), or to be honest, that is, truthfully return the correct result. If the master verifies, it then penalizes the cheaters and rewards the honest workers. Also, based on the number of cheaters, it might decide to increase or decrease the probability of verification for the next round.
If the master does not verify, then it accepts the result returned by the majority of the workers and rewards this majority (it does not penalize the minority); in this case, it does not change its probability of verifying. 
Similarly, depending on the payoff received in a given round (reward or punishment minus the cost of performing the task, if it 
performed it), each worker decides whether to increase or decrease its probability of cheating. It was shown that, under certain
conditions, eventually workers stop cheating and the master always obtains the correct task results with minimal verification. 

\noindent{\em Our approach.}
In this work, we take a different approach in exploiting the repeated interactions between the master and the rational workers.
We model this repeated interaction as a {\em repeated game}~\cite{osborne2}. Unlike the work in~\cite{ccpe}, as long as the
workers operate within this framework, the master obtains the correct task results (with high probability) from the very first round. 
The main idea is the following: when the workers detect that one worker (or more) has deviated from an agreed strategic choice, 
then they change their strategy into the one that maximizes the negative effect they have on the utility of the deviating worker.
This might negatively affect their own utility as well, but in long-running computations (such as master-worker computations) this 
punishment threat stops workers from deviating from the agreed strategic choice. So, indeed, workers do not deviate. (For more details on the theory of repeated games please refer to~\cite{osborne2}.)
As we demonstrate later, under certain conditions, not only the master obtains the correct results, but in the long run, it does so with lower cost when compared with the repeated use of the one-shot mechanism of~\cite{plosone} or the reinforcement learning mechanism of~\cite{ccpe}. 

\noindent{\em Contribution.}
In summary, the main contributions of this work are the following:\vspace{-.2em}
\begin{itemize}\itemsep2pt
\item To the best of our knowledge, this is the first work that attempts to increase the reliability of Internet-based master-worker computing by modeling the repeated interaction between the master and the workers as a repeated game. (The model is formalized in Section~\ref{sec:model}.)
\item We first present a mechanism (Section~\ref{section:pure}) where workers decide to cheat or be honest deterministically (in game-theoretic terms they follow pure
equilibria strategies), and prove the conditions and the cost under which the master obtains the correct task result in every round (with high probability). In order to allow the workers to detect other workers' deviations (from the agreed strategy), the master needs to provide only the number of different answers received (regardless of whether it verified or not). 
\item Then, we consider the case where the workers' decision is probabilistic (in game-theoretic terms they follow mixed equilibria strategies).
The mechanism (Section~\ref{section:mixed}) in this case is more involved as workers need more information in order to detect deviations. The master provides the workers
which answers it has received and how many of each, and the workers use this information to detect deviations. We prove the conditions
and the cost under which the master obtains the correct task result in every round (with high probability).
\item Finally, we perform a simulation study to demonstrate the utility of our new approach (Section~\ref{sec:sims}). The study complements the theoretical analysis
by providing more insight on the effectiveness of the mechanisms by experimenting on various parameter values, and also
provides comparison with the works in~\cite{plosone} and~\cite{ccpe}. In particular, our simulations show that in the presence of 
$\lceil n/2 \rceil$ deviators (out of $n$ total workers), which is the minimum to have an impact in voting mechanisms,
our mechanism performs similarly or better than the reinforcement learning mechanism of~\cite{ccpe}, and both mechanisms perform significantly better than the repeated use of the one-shot mechanism of~\cite{plosone}.   
\end{itemize}     

\noindent{\em Other related work.}
A classical Distributing Computing approach for increasing the reliability of master-working computing is 
to model the malfunctioning (due to a  hardware or a software error) or cheating (intentional wrongdoer) 
as {\em malicious} workers that wish to hamper the computation and thus always return an incorrect result. The
non-faulty workers are viewed as {\em altruistic} ones~\cite{boinc} that always return the correct result. 
Under this view, malicious-tolerant protocols have been considered, e.g.,~\cite{Sarmenta02, ALEX, PPL12}, 
where the master decides on the correct result based on majority voting. More recent works, e.g.,~\cite{IEEETC14,OPODIS13}, 
have combined this approach with incentive-based game theoretic approaches and devised mechanisms assuming the 
co-existence of altruistic, malicious and rational workers. The work in~\cite{OPODIS13} employed worker {\em reputation}
to cope with malice, while the work in~\cite{IEEETC14} relied on statistical information on the distribution of the worker
types (altruistic, malicious or rational). Extending our present work to cope with malicious workers is an interesting 
future direction. 
\remove{
- I think we do not need to mention the work with collisions here, we could just mention in the conclusion section
that another extension would be to consider worker collusion, as we did in the PLOSone paper. Is this enough?\\
- Have one paragraph presenting our evolutionary work and how it is different from this work. Beside the clear
difference mentioned above, that the master obtains the correct result after convergence is achieved, which
might take some rounds, I believe another disadvantage is that you need to also deal with the workers aspirations.
And the master needs to cover not only the cost of the task (like this work) but also the aspiration of the worker.
If workers report higher aspirations, the computation is becoming unnecessarily more expensive than needed (right?).\\
- Should we present more general related work with respect to repeated games? Are we familiar with such works?
}

\section{Model}
\label{sec:model}


\remove{
\mig{put here the details of the model such as payoffs, verification, etc.}
To achieve larger utility than in~\cite{FGM:mechdesgnConf}, we consider here that the penalty applied to the cheaters when verifying is proportional to the number of cheaters. \marginpar{MM: I do not remember why we decided this.}
\mig{complete}
}

\noindent{\em The master-workers framework.}
We consider a distributed system consisting of a master processor that assigns, over the Internet, 
computational tasks to a set $W$ of $n$ workers. In particular, the computation is broken into
rounds. In each round the master sends a task to be computed to the workers and 
the workers return the task  result. The master, based on the workers' replies, must decide on the value 
it believes is the correct outcome of the task in the same round. 
The tasks considered in this work are assumed to have a 
unique solution; although such limitation reduces the scope of application of the presented mechanism \cite{TACB05}, 
there are plenty of computations where the correct solution is unique: e.g., any mathematical function.
In this work security issues are not considered. Security can be achieved by cryptographic means, as done in  BOINC~\cite{boinc}, which allows for encrypting communication, authenticating master and workers, signing the code of tasks, and executing tasks in sandboxes. 

Following~\cite{Halp06} and~\cite{rational}, we consider workers to be {\em rational}, that is, they are selfish in a game-theoretic
sense and their aim is to maximize their benefit (utility) under 
the assumption that other workers do the same. In the context of this paper, a worker is {\em honest} 
in a round when it truthfully computes and returns the task result, and it {\em cheats} when it returns some 
incorrect value. We denote by $p^{r}_{Ci}$ the probability of a worker $i$ cheating in round $r$. 
Note that we do not consider non-intentional errors produced by hardware or software problems.

\remove{
\cg{Correct me if I am wrong, but I believe we do not need the assumption described below in this work, right?
[[While it is assumed that workers make their decision individually and with no coordination, it is assumed that all the workers that cheat in a round return the same incorrect value (as done, for example, in \cite{Sarmenta02,ccpe}).
This assumption yields a worst case scenario (and hence analysis) for the master
with respect to obtaining the correct result; 
it subsumes models where cheaters do not 
necessarily return the same answer. (In some sense, this can be seen as a cost-free, weak form of collusion.)]]}  
}

To ``enforce'' workers to be honest, the master employs, when necessary, {\em verification} and
{\em reward/punish} schemes. The master, in a round, might decide to verify the response of the workers, at a cost. 
It is assumed that verifying an answer is more efficient than computing the task~\cite{GS:secureDC} (e.g., $NP$-complete problems if $P\neq NP$), but the correct result of the computation is not obtained if the verification fails (e.g., when all workers cheat).
We denote by $p_\VRF$ the probability of the master verifying the responses of the workers. The goal of the master is to 
accept the correct task result in every round, while reducing its utility; therefore, verification needs to be used only when it is necessary.   
Furthermore, the master can reward and punish workers using the following scheme: When the master verifies, 
it can accurately reward and punish workers. When the master does not verify, it decides on the majority of the received 
replies, and it rewards only the majority (and it does not penalize the minority); probability is used to break symmetry.
This is essentially the reward model \modelone on the game \gamethrees as defined in~\cite{FGM:mechdesgnConf} (also
considered in other works, e.g.,~\cite{ccpe,IEEETC14}).
     
The payoff parameters considered in this work are detailed in Table~\ref{table:payoffs}.
Observe that there are different parameters for the reward 
$\Wb$ to a worker and the cost $\Ac$ of this reward to the master. This models the fact
that the cost to the master might be different from the benefit for a
worker.

\begin{table}[h]\centering
\begin{small}
\begin{tabular}{|c|l|}
\hline
$\Wp$& worker's penalty for being caught cheating\\
\hline
$\Wc$& worker's cost for computing the task\\
\hline
$\Wb$& worker's benefit from master's acceptance\\
\hline
$\Mp$& master's penalty for accepting a wrong answer\\
\hline
$\Ac$& master's cost for accepting the worker's answer\\
\hline
$\Vc$& master's cost for auditing worker's answers\\
\hline
$\Mb$& master's benefit from accepting the right answer\\
\hline
\end{tabular}
\vspace{.05in}
\caption{Payoffs. The parameters are non-negative.}
\label{table:payoffs}
\end{small}
\end{table}

For the purposes of repeated game framework (presented next), the punishment of the master to a worker caught cheating 
is proportional to the number of cheaters: Let $F$ denote the set of workers caught cheating in a round that the master
verifies. Then, in that round, the master applies penalty $\Wp\cdot|F|$ to {\em every} worker in $F$. 
The fact that punishment is proportional to the number of cheaters is, intuitively, an important tool to implement peer-punishment, which is required in the repeated games framework. Hence, we carry our analysis for proportional punishments. The study of constant punishments is an open question that we leave for future work.
\remove{
Among the parameters involved, we assume that the 
master has the freedom of choosing $\Wb$ and $\Wp$; all other parameters can either be 
fixed because they are system parameters, or may also be chosen by the master.}

\noindent{\em The repeated game framework.}
We assume that workers participate in the system within the framework of a \emph{repeated game}~\cite{osborne2}. The objective of the repeated games model is to capture the effect of long-term interaction. That is, workers take into account that their behavior in one computation will have an effect on the behavior of other workers in the future. We further assume that workers behave according to the reality that they perceive. That is, although their participation is physically bounded to a finite number of rounds, it is known~\cite{osborne2} that workers participate in the game as if the number of repetitions is infinite, even for a small number of repetitions, until the last repetition is close.
(Note that prediction mechanisms such as the one in~\cite{Kondopredict}, can be used to establish the availability of a set of workers for a sufficiently long period of time.) In this paper, we assume that workers participate unaware of when their participation will end, and we analyze the system as an \emph{infinite repeated game}. In the following game specification, we follow the notation in~\cite{osborne2}. 

Let the set of workers be $W$, and for each worker $i\in W$ let the set of strategies cheat and not-cheat be $A_i=\{\CH,\NCH\}$, and the preference relation (which is the obvious preference for a higher worker utility) be $u_i:A\to \mathbb{R}, A=\times_{j\in W}A_j$. Then, we model each round of computation as the mixed extension $G'=\langle W, (\Delta(A_i))_{i\in W}, (U_i)_{i\in W}\rangle$ of the strategic game $G=\langle W, (A_i)_{i\in W}, (u_i)_{i\in W}\rangle$, where $\Delta(A_i)$ is the set of probability distributions over $A_i$, and $U_i:\times_{j\in W}\Delta(A_j) \to \mathbb{R}$ is the expected utility of worker $i$ under $u_i$ induced by $\times_{j\in W}\Delta(A_j)$.

On the other hand, we model the multiple-round long-running computation as an infinitely repeated game where the constituent game is $G'$. That is, an extensive game with simultaneous moves $\langle W,H,P,(U_i^*)_{i\in W}\rangle$, where $H=\{\emptyset\}\cup(\cup_{r=1}^\infty A^r)\cup A^\infty$, $\emptyset$ is the initial empty history, $A^\infty$ is the set of infinite sequences $(a^r)_{r=1}^\infty$ of action profiles in $G$, the (next) player function is $P(h)=W$ (simultaneous moves) for each history $h\in H$, and $U_i^*$ is a preference relation on $A^\infty$ that extends the preference relation $U_i$ to infinite rounds under the limit of means criterion. That is, the payoff of all rounds is evaluated symmetrically, in contrast with other criteria where the value of a given gain may diminish with time. 

We are interested in equilibria that harness long-term objectives of the workers, rather than strategies that apply to short-sighted workers that would isolate each round of computation as a single game.
To support equilibria of the infinitely repeated game that are not simple repetitions of equilibria in the constituent game, workers are deterred from deviating by being punished by other workers. Specifically, any deviation from an agreed equilibrium, called a \emph{trigger strategy}, of some worker $i$ is punished by all other workers changing their strategy to enforce the \emph{minmax payoff} of $v_i$, that is, the lowest payoff that the other workers can force upon worker $i$. 

Any payoff profile $w$, where for each worker it is $w_i\geq v_i$, is called \emph{enforceable}. The Nash folk theorem for the limit of means criterion~\cite{osborne2} establishes that every feasible\footnote{Any convex combination of payoff profiles of outcomes in $A$} enforceable payoff profile of the constituent game is a Nash equilibrium payoff profile of the limit of means infinitely repeated game. Thus, any trigger strategy that yields an enforceable payoff profile is an equilibrium. So, we focus our effort in finding the minmax payoff to later analyze which of the infinitely many trigger criteria yields a mechanism that is beneficial for the master and the workers.
For clarity, we present our mechanisms punishing the deviant indefinitely. Nevertheless, it is enough to held the deviant's payoff to the minmax level for enough rounds to wipe out its gain from the deviation, as shown by the Perfect folk theorem~\cite{osborne2}.  

In order to make punishment decisions, workers need information about previous outcomes. We consider two scenarios, one where the master only provides the number of different answers received, and another where the master informs which are the answers received, and how many of each. In the first case, workers are bounded to use pure equilibria of the constituent game, so that non-deviators can decide to punish if there were workers replying with an answer other than theirs. 
For the second case, a worker that did not cheat may count the number of answers different from its own. But a worker that cheated needs to know which one was correct. We assume that workers can verify the answers with negligible cost. Notice that the same does not apply to the master, which is assumed to have a high operation cost yielding a high cost of verification.

\section{Limited Information (pure equilibria)}
\label{section:pure}

In this section we assume that workers decide to be honest or to cheat deterministically, i.e., follow pure strategies. Under this assumption we define the mechanism shown in Algorithm~\ref{algpw}. We will show using repeated-games analysis that this mechanism leads to an equilibrium where no worker cheats. The algorithm defined requires in every round the number of different answers received by the master in the round. This is provided by the master, as shown in Algorithm~\ref{algpm}.


\begin{algorithm}[h]
\label{algpm}
\begin{small}
\caption{Pure strategies master algorithm. The probability of verification $p_{\VRF}$ is any value such that $\Wc/(\Wb + \Wp\lceil n/2\rceil) < p_\VRF < (\Wb+\Wc)/(2\Wb + n\Wp)$}
\SetKwFor{Upon}{upon}{do}{endupon}
\DontPrintSemicolon
\While{true}{
send a computational task to all the workers in $W$\;
\Upon{receiving all answers}{
 with probability $p_{\VRF}$, verify the answers\;\label{step:chp}
\lIf{the answers were not verified}{accept the majority}
reward/penalize accordingly\;
send to the workers the number of different answers received\;
}
}
\end{small}
\end{algorithm}

\begin{algorithm}[h]
\label{algpw}
\begin{small}
\caption{Pure strategies worker algorithm. 
}
\SetKwFor{Upon}{upon}{do}{endupon}
\DontPrintSemicolon
strategy $\leftarrow \NCH$\; 
\While{true}{
\Upon{receiving a task}{
\If{strategy$=\NCH$}{
compute the task and send the result to the master\;
}
\Else{
do not compute and send a bogus result to the master\;
}
\Upon{receiving from the master the number of different answers}{
\If{number of different answers $>1$}{
strategy $\leftarrow \CH$\;
}
}
}
}
\end{small}
\end{algorithm}

We analyze now the properties of this mechanism. Although the algorithms allow that a cheating worker replies with any value, in the analysis of the rest of the section we assume that 
all workers that cheat in a round return the same incorrect value (as done also, for example, in \cite{Sarmenta02,ccpe}). 
Intuitively, this assumption yields a worst case scenario (and hence analysis) for the master
with respect to obtaining the correct result.

Recall that, for any given round of the constituent game, $U_i$ is the expected utility of worker $i$ and $s_i$ is the strategy chosen by worker $i$. Let $F$ be the set of cheaters and $F_{-i}=\{j|j\in W\land j\neq i \land s_j=\CH\}$. 
Then, the following holds.

\begin{lemma}
\label{lemma:pureminmax}
If $\Wb>\Wc$ and $\Wp>\Wc$,
for any $p_\VRF$ such that $\Wc/(\Wb + \Wp\lceil n/2\rceil) < p_\VRF < (\Wb+\Wc)/(2\Wb + n\Wp)$,
and for any worker $i\in W$,
the minmax expected payoff is $v_i = (1-p_\VRF) \Wb- p_\VRF n \Wp$, 
which is obtained when $s_i=\CH$ and $|F_{-i}|>(n-1)/2$. 
\end{lemma}
\begin{IEEEproof}
We notice first that, given that $\Wb>\Wc$ and $\Wp>\Wc$, the range of values $\Wc/(\Wb + \Wp\lceil n/2\rceil) < p_\VRF < (\Wb+\Wc)/(2\Wb + n\Wp)$ for $p_\VRF$ is not empty.
%
For any worker $i$, there are four possible utility outcomes, namely.
\begin{align}
U_i(s_i=\CH,|F|>n/2) &=  (1-p_\VRF) \Wb- p_\VRF \Wp|F|\label{allcheat}\\
U_i(s_i=\NCH,|F|>n/2) &= p_\VRF \Wb-\Wc\label{onehonest}\\
U_i(s_i=\CH,|F|<n/2) &= - p_\VRF \Wp|F|\label{onecheat}\\ 
U_i(s_i=\NCH,|F|<n/2) &= \Wb-\Wc\label{allhonest} 
\end{align}

We want to find worker $i$'s minmax payoff, which is the lowest payoff that the other workers can force upon $i$ (cf.~\cite{osborne2}). That is,
\begin{align*}
v_i = \min_{s_{-i}\in \{\CH,\NCH\}^{n-1}} \max_{s_i\in \{\CH,\NCH\}} U_i(s_i,s_{-i}).
\end{align*}

From the perspective of worker $i$, the actions of the remaining workers fall in one of three cases: $|F_{-i}|< (n-1)/2$, $|F_{-i}|= (n-1)/2$, and $|F_{-i}|> (n-1)/2$.
Thus, we have 
\begin{multline}
v_i=\min\{
\max_{s_i\in \{\CH,\NCH\}} U_i(s_i,|F_{-i}|<(n-1)/2),
\max_{s_i\in \{\CH,\NCH\}} U_i(s_i,|F_{-i}|=(n-1)/2),\\
\max_{s_i\in \{\CH,\NCH\}} U_i(s_i,|F_{-i}|>(n-1)/2)
\}\label{minmax}
\end{multline}

From Equations~\ref{onecheat} and~\ref{allhonest},
\begin{multline*}
\max_{s_i\in \{\CH,\NCH\}} U_i(s_i,|F_{-i}|<(n-1)/2) 
= \max\{- p_\VRF \Wp|F|,\Wb-\Wc\}, \textrm{ for $1\leq |F|<\lceil n/2\rceil$.}
\end{multline*}

Given that $\Wb>\Wc$, it is
\begin{align}
\max_{s_i\in \{\CH,\NCH\}} U_i(s_i,|F_{-i}|<(n-1)/2) &= \Wb-\Wc.\label{othershonest}
\end{align}

From Equations~\ref{allcheat} and~\ref{onehonest},
\begin{multline*}
\max_{s_i\in \{\CH,\NCH\}} U_i(s_i,|F_{-i}|>(n-1)/2) 
= \max\{ (1-p_\VRF) \Wb- p_\VRF \Wp|F|, p_\VRF \Wb-\Wc\}, \textrm{ for $\lceil n/2\rceil<|F|\leq n$.}
\end{multline*}


Given that $p_\VRF < (\Wb+\Wc)/(2\Wb + n\Wp)$, it is
\begin{align}
\max_{s_i\in \{\CH,\NCH\}} U_i(s_i,|F_{-i}|>(n-1)/2) 
&= (1-p_\VRF) \Wb- p_\VRF \Wp|F|, \textrm{ for $\lceil n/2\rceil<|F|\leq n$.}\label{otherscheat}
\end{align}

From Equations~\ref{allcheat} and~\ref{allhonest},
\begin{multline*}
\max_{s_i\in \{\CH,\NCH\}} U_i(s_i,|F_{-i}|=(n-1)/2)
= \max\{ (1-p_\VRF) \Wb- p_\VRF \Wp\lceil n/2\rceil, \Wb-\Wc \}.
\end{multline*}


Given that $p_\VRF>\Wc/(\Wb + \Wp\lceil n/2\rceil)$, it is
\begin{align}
\max_{s_i\in \{\CH,\NCH\}} U_i(s_i,|F_{-i}|=(n-1)/2)
&= \Wb-\Wc.\label{othersequal}
\end{align}

Replacing Equations~\ref{othershonest},~\ref{otherscheat}, and~\ref{othersequal} in Equation~\ref{minmax}, we obtain
\begin{align*}
v_i
&= \min\{\Wb-\Wc,(1-p_\VRF) \Wb- p_\VRF \Wp|F|\}, &\textrm{ for $\lceil n/2\rceil<|F|\leq n$}\\
&= (1-p_\VRF) \Wb- p_\VRF n \Wp.
\end{align*}


The latter is true because $p_\VRF>\Wc/(\Wb + \Wp\lceil n/2\rceil)>\Wc/(\Wb + n\Wp)$.
Thus, the claim holds.
\end{IEEEproof}

The following theorem establishes the correctness of our pure-strategies mechanism. The proof follows from Lemma~\ref{lemma:pureminmax} and the repeated-games framework~\cite{osborne2}.
\begin{theorem}
\label{thm:pure}
For a long-running multi-round computation system with set of workers $W$, 
and for a set of payoff parameters such that $\Wb>\Wc$ and $\Wp>\Wc$,
the mechanism defined in Algorithms~\ref{algpm} and~\ref{algpw} guarantees that the master always obtains the correct answer. In each round, the utility of each worker $i$ is $U_i=\Wb-\Wc$ and the expected utility of the master is $U_M=\Mb-n\Ac-p_{\VRF}\Vc$.
\end{theorem}

\section{More Information Helps (mixed equilibria)}
\label{section:mixed}


In this section, we consider the general case in which workers can randomize their decision to be honest of cheat. In this case, the equilibrium in the constituent game is mixed, that is, in the equilibrium workers cheat with some probability $p \in (0,1)$, rather than behaving deterministically.
Hence, the actual probability used by each worker cannot be inferred accurately from the outcome of one computation.
In other words, even knowing that some given worker has cheated after a computation, it is not possible to know if this event was a deviation 
from the equilibrium or not from one single worker outcome.
Nonetheless, it is possible to provide stochastic guarantees, either from many computations of one worker, or one computation by multiple workers. Such guarantees may be enough for some scenarios.

Specifically, if the master announces how many of each answer has received, workers may make punishment decisions based on the probability of such outcome. This is possible even for workers that did not compute the task, given that the cost of verification for workers is assumed to be negligible. Such decision will not be based on deterministic information, but we can provide guarantees on the probability of being the right decision. Hence, in this section we define a mechanism where the
master sends to all the workers in each round the answers that it has received and how many of each. This is described in Algorithm~\ref{algmm}.
%
%
In summary, the approach is to carry out the computation as in a regular repeated game, but punishments are applied when it is known that some workers have deviated from the equilibrium of the constituent game with some parametric probability. 
The mechanism that is assigned to the workers is described in Algorithm~\ref{algmw}. 

We emphasize that what is punished by peer-workers is deviation from the agreed equilibrium (rather than cheating which is punished by the master). That is, if the equilibrium is to cheat with some probability $p$ and the number of incorrect answers was $x$, punishment is applied when the probability of $x$ incorrect answers is very low if all workers were using $p$, either if $x$ is less or more than the number of cheaters expected.

It is known~\cite{osborne2} that it is enough to apply peer-punishment for a number of rounds that neutralizes the benefit that the deviators might have obtained by deviating. Nonetheless, to avoid unnecessary clutter, we omit this detail in Algorithm~\ref{algmw} where punishment is applied forever. In our simulations in Section~\ref{sec:sims}, we limit the punishment to one round, since that is enough to neutralize the benefit of a deviation for those parameters.


\begin{algorithm}[h]
\label{algmm}
\begin{small}
\caption{Mixed strategies master algorithm. $p_\VRF$ is set according to Lemma~\ref{lemma:chernoffcorrect}.}
\SetKwData{answer}{answer}
\SetKwData{count}{count}
\SetKwFor{Upon}{upon}{do}{endupon}
\DontPrintSemicolon
\While{true}{
send a computational task to all the workers in $W$\;
\Upon{receiving all answers}{
 with probability $p_{\VRF}$, verify the answers\;\label{step:chm}
\lIf{the answers were not verified}{accept the majority}
reward/penalize accordingly\;
send to the workers a list of pairs $\langle \answer, \count\rangle$\;
}
}
\end{small}
\end{algorithm}

\begin{algorithm}[h]
\label{algmw}
\begin{small}
\caption{Mixed strategies worker algorithm.  
$p_\CH$ is initialized according to Lemma~\ref{lemma:chernoffcorrect}.
$\varepsilon>0$ is the probability of erroneous punishment (cf. Lemma~\ref{lemma:chernoffpunish}).
}
\SetKwData{L}{counts}
\SetKwData{correct}{correct}
\SetKwData{incorrect}{\#incorrect}
\SetKwData{answer}{answer}
\SetKwData{count}{count}
\SetKwData{result}{result}
\SetKwData{round}{round}
\SetKwData{cheat}{cheat}
\SetKwData{cheaters}{cheaters}
\SetKwData{maxr}{maxrounds}
\SetKwFor{Upon}{upon}{do}{endupon}
\DontPrintSemicolon
\maxr$\leftarrow \lfloor3 n p_\CH \ln(1/\varepsilon)\rfloor$
\tcp*{Punishment decisions only for $\delta\geq1/np_\CH$ (cf. Lemma~\ref{lemma:chernoffpunish}).}
$\L \leftarrow$ empty queue of integers
\tcp*{$\L[i]$ is the $(i+1)$th item, for $i=0,1,2,\dots$.}
\For{ each $\round=1,2,\dots$}{
	\Upon{receiving a task}{
		\tcp{computation phase}
		$
                \text{\cheat} \leftarrow
                \begin{cases}
			true, & \text{with probability } p_{\CH}\\
			false, & \text{with probability } 1-p_{\CH}\\
                \end{cases}
		$\;
		\lIf{$\cheat=false$}{$\result\leftarrow$ task result computed}
		\lElse{$\result\leftarrow$ bogus result}
		send $\result$ to the master\;
		\tcp{punishment phase}
		\Upon{receiving from the master a list of pairs $\langle \answer, \count\rangle$}{
			\tcp{update \# of cheaters per round}
			verify all answers\; 
			$\incorrect\leftarrow$ number of incorrect answers\;
			enqueue $\incorrect$ to $\L$\;
			\lIf{size of $\L>\maxr$}{dequeue from $\L$}
			\tcp{punishment decision}
			$\cheaters_{\min} \leftarrow n$\;
			$\cheaters_{\max} \leftarrow 0$\;
			$R\leftarrow\min\{\maxr,\round\}$\;
			\For{$r=1$ to $R$}{
				\lIf{$\L[R-r]<\cheaters_{\min}$}{$\cheaters_{\min} \leftarrow \L[R-r]$}	
				\lIf{$\L[R-r]>\cheaters_{\max}$}{$\cheaters_{\max} \leftarrow \L[R-r]$}
				$\delta\leftarrow\sqrt{3\ln (1/\varepsilon)/(r n p_\CH)}$\;
				\If{$\delta<1$}{
					\If(\tcp*[f]{Lemma~\ref{lemma:chernoffpunish}}){
							$\cheaters_{\min}\geq\lceil(1+ \delta)np_\CH\rceil$ {\bf or}
							$\cheaters_{\max}\leq\lfloor(1- \delta)np_\CH\rfloor$ 
					}{
						$p_{\CH}\leftarrow 1$
						\tcp*[f]{Lemma~\ref{lemma:mixedminmax}}
					}
				}
			}
		}
	}
}
\end{small}
\end{algorithm}

As in the analysis of the previous section, to obtain a worse case analysis for the master, we assume that cheating workers return the same incorrect result.


The following lemma characterizes the number of incorrect answers and the number of rounds that should trigger a punishment.
\begin{lemma}
\label{lemma:chernoffpunish}
In a system with $n$ workers where the mixed equilibrium of the constituent game is to cheat with some probability $p_{\CH}>0$, 
if the number of incorrect answers is at least $(1+\delta)\lceil np_{\CH}\rceil$ or at most $(1-\delta)\lfloor np_{\CH}\rfloor$ during $r$ consecutive rounds, 
for any $1/(np_\CH)\leq \delta<1$, $r\geq 1$, and $\varepsilon>0$ such that $r\delta^2>3 \ln (1/\varepsilon)/(np_{\CH})$, then there are one or more workers deviating from that equilibrium with probability at least $1-\varepsilon$.
\end{lemma}
\begin{IEEEproof}
Let $1,2,\dots,n$ be some labeling identifying the workers.
For any given round of computation, let $X_i$ be a random variable indicating whether worker $i$ cheated or not, and let $X=\sum_{i=1}^n X_i$.
For $i=1,2,\dots,n$, the $X_i$ random variables are not correlated. 
Thus, we can upper bound the tails of the probability distribution on the number of cheaters using the following Chernoff-Hoeffding bounds~\cite{book:mitzenmacher}.
For $0<\delta\leq 1$, it is $Pr(X\geq (1+\delta)n p_{\CH}) \leq e^{-n p_{\CH}\delta^2/3}$ 
and for $0<\delta< 1$, it is $Pr(X\leq (1-\delta)n p_{\CH}) \leq e^{-n p_{\CH}\delta^2/2}$.
Therefore, it is $Pr(X\geq (1+\delta)\lceil n p_{\CH}\rceil) \leq e^{-n p_{\CH}\delta^2/3}$ 
and $Pr(X\leq (1-\delta)\lfloor n p_{\CH}\rfloor) \leq e^{-n p_{\CH}\delta^2/2}$. Given that $X$ cannot differ  from the expected number of cheaters by less than one worker, we further restrict $\delta$ as follows.
For $1/(n p_{\CH})\leq\delta< 1$, it is $Pr(X\geq (1+\delta)\lceil n p_{\CH}\rceil) \leq e^{-n p_{\CH}\delta^2/3}$ 
and $Pr(X\leq (1-\delta)\lfloor n p_{\CH}\rfloor) \leq e^{-n p_{\CH}\delta^2/2}$.

Letting $E_{high}$ be the event of having $X\geq (1+\delta)\lceil n p_{\CH}\rceil$ 
incorrect answers for $r$ consecutive rounds,
and $E_{low}$ be the event of having $X\leq (1-\delta)\lfloor n p_{\CH}\rfloor$ 
incorrect answers for $r$ consecutive rounds,
it is $Pr(E_{high}) \leq e^{-n r p_{\CH}\delta^2/3}$ and
it is $Pr(E_{low}) \leq e^{-n r p_{\CH}\delta^2/2}$.
Given that $e^{-n r p_{\CH}\delta^2/2} \leq e^{-n r p_{\CH}\delta^2/3} \leq \varepsilon$ for $r\delta^2 \geq 3\ln (1/\varepsilon)/(n p_{\CH})$, 
if either $E_{high}$ or $E_{low}$ occurs, there are one or more workers deviating from equilibrium with probability at least $1-\varepsilon$ and the claim follows.
\end{IEEEproof}


In the following lemma, we show what is the minmax payoff when a mixed equilibrium of the constituent game is allowed.

\begin{lemma}
\label{lemma:mixedminmax}
For any $p_\VRF$, such that 
$p_\VRF > 2\Wb/(2\Wb+\Wp n)$
and $p_\VRF \geq\Wc/\Wb$,
and for any worker $i\in W$,
the minmax expected payoff of worker $i$ is $v_i=p_\VRF \Wb-\Wc$, which is attained when all the other workers use $p_{\CH}=1$. 
\end{lemma}

\begin{IEEEproof}
Let $\sigma$ be a mixed strategy profile, that is, a mapping from workers to probability distributions over the pure strategies $\{\CH,\NCH\}$, 
let $\sigma_i$ be the probability distribution $\{p_{\CH_i},(1-p_{\CH_i})\}$ of worker $i$ in $\sigma$,
and let $\sigma_{-i}$ be the mixed strategy profile of all workers but $i$.
We want to find worker $i$'s minmax payoff, which is the lowest payoff that the other workers can force upon $i$ (cf.~\cite{osborne2}). That is,
\begin{align*}
v_i = \min_{\sigma_{-i}} \max_{\sigma_i} U_i(\sigma_i,\sigma_{-i}).
\end{align*}

For any worker $i$, there are four possible utility outcomes, namely.
\begin{align*}
U_i(s_i=\CH,|F|>n/2) &=  (1-p_\VRF) \Wb- p_\VRF \Wp|F|\\
U_i(s_i=\NCH,|F|>n/2) &= p_\VRF \Wb-\Wc\\
U_i(s_i=\CH,|F|<n/2) &= - p_\VRF \Wp|F|\\ 
U_i(s_i=\NCH,|F|<n/2) &= \Wb-\Wc. 
\end{align*}

The expected utility of a worker $i$ that deviates from equilibrium when all other workers use the same mixed strategy $p_{\CH_{-i}}$ is the following.

\begin{align*}
U_i 
&= p_{\CH_i} (Pr\left(|F_{-i}|\geq\frac{n-1}{2}\right) U_i\left(s_i=\CH,|F|>\frac{n}{2}\right) 
+ Pr\left(|F_{-i}|<\frac{n-1}{2}\right) U_i\left(s_i=\CH,|F|<\frac{n}{2}\right))\\
&+(1-p_{\CH_i}) (Pr\left(|F_{-i}|>\frac{n-1}{2}\right) U_i\left(s_i=\NCH,|F|>\frac{n}{2}\right) 
+ Pr\left(|F_{-i}|\leq\frac{n-1}{2}\right) U_i\left(s_i=\NCH,|F|<\frac{n}{2}\right)).
\end{align*}

It can be seen that $U_i$ is linear with respect to $p_{\CH_i}$. That is, 
the function is either monotonically increasing, monotonically decreasing, or constant with respect to $p_{\CH_i}$, depending on the relation among the parameters, but it does not have critical points. Hence, for any given $p_{\CH_{-i}}$ the maximum utility for worker $i$ occurs when $p_{\CH_i}$ is either $0$ or $1$. We get then that the maximum $U_i$ is either
\begin{align*}
U_i (p_{\CH_i}=1)
&= Pr\left(|F_{-i}|\geq\frac{n-1}{2}\right) U_i\left(s_i=\CH,|F|>\frac{n}{2}\right) 
+ Pr\left(|F_{-i}|<\frac{n-1}{2}\right) U_i\left(s_i=\CH,|F|<\frac{n}{2}\right), \textrm{ or}\\
U_i (p_{\CH_i}=0)
&= Pr\left(|F_{-i}|>\frac{n-1}{2}\right) U_i\left(s_i=\NCH,|F|>\frac{n}{2}\right) 
+ Pr\left(|F_{-i}|\leq\frac{n-1}{2}\right) U_i\left(s_i=\NCH,|F|<\frac{n}{2}\right).
\end{align*}
Replacing and using that $\sum_{j=0}^{n-1} \binom{n-1}{j} p_{\CH_{-i}}^{j} (1-p_{\CH_{-i}})^{n-1-j} =1$, and that $\sum_{j=0}^{n-1} \binom{n-1}{j} p_{\CH_{-i}}^{j} (1-p_{\CH_{-i}})^{n-1-j}j=(n-1)p_{\CH_{-i}}$ we have the following.
\begin{align}
&U_i (p_{\CH_i}=1)
=  \sum_{j=(n-1)/2}^{n-1} \binom{n-1}{j} p_{\CH_{-i}}^j (1-p_{\CH_{-i}})^{n-1-j}
\cdot((1-p_\VRF) \Wb- p_\VRF \Wp(j+1))\nonumber\\
&+ \sum_{j=0}^{(n-3)/2} \binom{n-1}{j} p_{\CH_{-i}}^{j} (1-p_{\CH_{-i}})^{n-1-j}(- p_\VRF \Wp (j+1))\nonumber\\
&=  - p_\VRF \Wp(1+(n-1)p_{\CH_{-i}}) 
+ (1-p_\VRF) \Wb\sum_{j=(n-1)/2}^{n-1} \binom{n-1}{j} p_{\CH_{-i}}^j (1-p_{\CH_{-i}})^{n-1-j}.\label{2ndmax}
\end{align}
And
\begin{align}
&U_i (p_{\CH_i}=0)
= \sum_{j=(n+1)/2}^{n-1} \binom{n-1}{j} p_{\CH_{-i}}^j (1-p_{\CH_{-i}})^{n-1-j}(p_\VRF \Wb-\Wc)\nonumber\\
&+ \sum_{j=0}^{(n-1)/2} \binom{n-1}{j} p_{\CH_{-i}}^j (1-p_{\CH_{-i}})^{n-1-j}(\Wb-\Wc)\nonumber\\
&= p_\VRF \Wb-\Wc
+ (1-p_\VRF)\Wb \sum_{j=0}^{(n-1)/2} \binom{n-1}{j} p_{\CH_{-i}}^j (1-p_{\CH_{-i}})^{n-1-j}.\label{1stmax}
\end{align}

To find the minmax payoff, we want to find the mixed strategy (i.e., $p_{\CH_{-i}}$) that other workers may choose to minimize these utilities.
Equation~\ref{1stmax} is minimized when $p_{\CH_{-i}}=1$, yielding $U_i(p_{\CH_i}=0,p_{\CH_{-i}}=1)=p_\VRF \Wb-\Wc\geq 0$. The latter inequality is true because $p_\VRF \geq\Wc/\Wb$.  
On the other hand, given that $p_\VRF > 2\Wb/(2\Wb+\Wp n)$, it is $p_\VRF \Wp n > 2(1-p_\VRF) \Wb$. Then, from Equation~\ref{2ndmax}, we have
\begin{align}
&U_i(p_{\CH_i}=1)
< - p_\VRF \Wp (1-p_{\CH_{-i}}) - 2(1-p_\VRF) \Wb p_{\CH_{-i}} 
+ (1-p_\VRF) \Wb \sum_{j=(n-1)/2}^{n-1} \binom{n-1}{j} p_{\CH_{-i}}^j (1-p_{\CH_{-i}})^{n-1-j}\nonumber\\
&= - p_\VRF \Wp (1-p_{\CH_{-i}}) +(1-p_\VRF) \Wb
\cdot\left(- 2 p_{\CH_{-i}} + \sum_{j=(n-1)/2}^{n-1} \binom{n-1}{j} p_{\CH_{-i}}^j (1-p_{\CH_{-i}})^{n-1-j}\right).\label{bound}
\end{align}
Given that $(- 2 p_{\CH_{-i}} + \sum_{j=(n-1)/2}^{n-1} \binom{n-1}{j} p_{\CH_{-i}}^j (1-p_{\CH_{-i}})^{n-1-j}) \leq 0$, Equation~\ref{bound} is negative for any $p_{\CH_{-i}}$. Therefore,  $U_i(p_{\CH_i}=0,p_{\CH_{-i}}=1)\geq U_i(p_{\CH_i}=1,p_{\CH_{-i}}=p)$ for any $p\in[0,1]$, and the minmax expected payoff is $v_i=U_i(p_{\CH_i}=0,p_{\CH_{-i}}=1)=p_\VRF \Wb-\Wc$.
\end{IEEEproof}

Given that the aim of the mechanism is to provide correctness, $p_{\VRF}$ must be lower bounded and $p_\CH$ upper bounded to enforce an equilibrium that provides correctness guarantees in each computation with parametric probability,
which we do in the following lemma.
\begin{lemma}
\label{lemma:chernoffcorrect}
In any given round of computation, 
and for any $\varphi>0$ and $0<\xi\leq 1$, 
if each worker cheats with probability $p_{\CH}<1/(2(1+\xi))$ and the master verifies with probability 
$p_\VRF \geq (e^{-n \xi^2/(6(1+\xi))} - \varphi)/(e^{-n \xi^2/(6(1+\xi))}-p_\CH^n)$,
the probability that the master obtains a wrong answer is at most $\varphi$.
\end{lemma}
\begin{IEEEproof}
Let $1,2,\dots,n$ be some labeling identifying the workers.
Let $X_i$ be a random variable indicating whether worker $i$ cheated or not, and let $X=\sum_{i=1}^n X_i$.
For $i=1,2,\dots,n$, the $X_i$ random variables are not correlated. 
Thus, we can upper bound the probability of having a majority of cheaters using the following Chernoff-Hoeffding bound~\cite{book:mitzenmacher}.
For $0<\xi\leq 1$, it is $Pr(X\geq (1+\xi)n p_{\CH}) \leq e^{-n p_{\CH}\xi^2/3}$. 
For $p_{\CH}<1/(2(1+\xi))$ we get 
\begin{align*}
Pr(X\geq (1+\xi)n p_{\CH}) &\leq e^{-n p_{\CH}\xi^2/3}\\
Pr(X> n/2) &\leq e^{-n \xi^2/(6(1+\xi))}.
\end{align*}
Thus, for the master to achieve correctness with probability at least $1-\varphi$, it is enough to have 
$(1-p_\VRF)e^{-n \xi^2/(6(1+\xi))} + p_\VRF p_\CH^n \leq \varphi$, which is true for 
%
%
%
$p_\VRF \geq \frac{e^{-n \xi^2/(6(1+\xi))} - \varphi}{e^{-n \xi^2/(6(1+\xi))}-p_\CH^n}$ if $p_\CH<1/e^{\xi^2/(6(1+\xi))}$, which holds for $p_{\CH}<1/(2(1+\xi))$.
\end{IEEEproof}


The following theorem establishes the correctness guarantees of our mixed-strategies mechanism. 
\begin{theorem}
\label{thm:mixed}
Consider a long-running multi-round computation system with set of workers such that $n>2$.
For any set of payoff parameters, 
and any $0<p_{\CH}<1/(2(1+\xi))$ for some $0<\xi\leq 1$,
setting 
the probability of verification so that
$p_\VRF > 2\Wb/(2\Wb+\Wp n)$,
$p_\VRF \geq\Wc/\Wb$,
and $p_\VRF\geq \frac{e^{-n \xi^2/(6(1+\xi))} - \varphi}{e^{-n \xi^2/(6(1+\xi))}-p_\CH^n}$, 
for some $\varphi>0$,
the following applies to each round of computation. 
If workers comply with the repeated games framework when punishment is stochastically consistent, 
the mechanism defined in Algorithms~\ref{algmm} and~\ref{algmw} guarantees that the master obtains the correct answer with probability at least $1-\varphi$, 
the expected utility of the master is 
\begin{align*}
U_M &= p_\VRF(\Mb-\Vc + n p_\CH\Wp - n(1-p_\CH)\Ac) 
+ (1-p_\VRF)(\sum_{i=0}^{\lfloor n/2\rfloor}\binom{n}{i}p_\CH^{i}(1-p_\CH)^{n-i}\left(\Mb - i\Ac\right)\\
&-\sum_{i=\lceil n/2\rceil}^{n}\binom{n}{i}p_\CH^i(1-p_\CH)^{n-i}\left(\Mp+ i\Ac\right)),
\end{align*}
and the expected utility of each worker $i$ is
\begin{align*}
U_i
&=  
\left(p_{\CH} (1-p_\VRF) p_{\geq h} 
+ (1-p_{\CH})p_\VRF  p_{> h}
+ (1-p_{\CH}) p_{\leq h}\right)\Wb
- p_{\CH}p_\VRF(1+(n-1)p_\CH)\Wp
- (1-p_\CH)\Wc.
\end{align*}
Where
\begin{align*}
p_{\geq h} &=  \sum_{j=(n-1)/2}^{n-1} \binom{n-1}{j} p_{\CH}^j (1-p_{\CH})^{n-1-j},\\
p_{> h} &=  \sum_{j=(n+1)/2}^{n-1} \binom{n-1}{j} p_{\CH}^j (1-p_{\CH})^{n-1-j}, \textrm{ and}\\
p_{\leq h} &=  \sum_{j=0}^{(n-1)/2} \binom{n-1}{j} p_{\CH}^j (1-p_{\CH})^{n-1-j}.
\end{align*}
\end{theorem}
\begin{IEEEproof}
First, we notice that, by making $p_\CH$ arbitrarily close to $0$, the expected worker utility is arbitrarily close to $\Wb-\Wc$, which is bigger than the minmax payoff $v_i=p_\VRF\Wb-\Wc$ for any $p_\VRF<1$. Then, there exists an enforceable payoff profile $w$, that is, a payoff profile where $w_i\geq v_i$ for each worker $i$ and, hence, there exists a Nash equilibrium payoff profile that all workers will follow due to long-term rationality (cf. Proposition 144.3 in~\cite{osborne2}).
The rest of the claim follows from Lemmas~\ref{lemma:chernoffpunish},~\ref{lemma:mixedminmax}, and~\ref{lemma:chernoffcorrect}. 
\end{IEEEproof}

\section{Simulations}
\label{sec:sims}

\subsection{Design}

In this section, we present our simulations of the mechanism in Algorithms~\ref{algmm} and~\ref{algmw}. For the sake of contrast, we also carry out simulations for the mechanism in~\cite{ccpe}, and for a repeated application of the one-shot mechanism in~\cite{plosone}. For easy reference, we denote our mechanism as \defn{RG} (repeated games), the mechanism in~\cite{ccpe} as \defn{ED} (evolutionary dynamics), the mechanism in~\cite{plosone} as \defn{OS} (one shot), and the repeated application of OS in a multi-round computation environment as \defn{ROS} (repeated one shot).

These mechanisms differ as follows. 
RG includes the threat of a peer-punishment that stops workers from deviating from the agreed equilibrium,
ED is aimed to converge to correctness after some time (rounds), 
and OS does not include previsions for equilibrium deviations.
%
The common assumption among all three approaches is that workers comply with the mechanism laid out. 
In RG, workers are assumed to be rational in the long term, and consequently they will agree on a given Nash Equilibrium (i.e., a $p_\CH$) that is computed taking into account the potential profit of future interactions.
In ED, it is assumed that workers update their $p_\CH$ for the next computation round using a particular formula, which is a function of the $p_\CH$ and the payment received in the previous computation round. 
In OS, workers are assumed to be rational but short-sighted, that is, they do not take into account future interactions (short-term rationality).

Once the parameters (payment, punishment, etc.) have been fixed (as we do for simulations), the performance of RG, ED, and ROS when all workers comply with the agreed protocol can be compared by a simple computation of expected utilities and probability of correctness. However, the tradeoffs between correctness and utilities when worker misbehavior occurs (in other words, what is the resilience of these systems to deviations) is an open question that we answer experimentally as follows.

For the sake of fair comparison, we use for all three mechanisms the same parameters used in~\cite{ccpe,plosone} whenever possible (summarized in Table~\ref{table:param} in the Appendix). Specifically, we set $\Wc =0.1$.
For each $n \in \{9, 27, 81\}$ and $\Wb \in \{1,1.1,1.2,\dots, 2\}$, we set $\Wp=\Wb/(np_\CH)$ for RG.
That is, the expected punishment in RG is $\Wb$.
Given that in ED and ROS the worker punishment is a constant value (i.e., it is not proportional to the number of cheaters), to simulate ED and OS we set $\Wp=\Wb$, that is, the expected punishment in RG.
As indicated in~\cite{ccpe}, these parameters are consistent with statistics obtained in SETI@home projects.

Observing the conditions of Theorem~\ref{thm:mixed}, for RG we fix 
$p_\CH=0.1<1/(2(1+\xi))$ for any $\xi>0$,
and we fix $p_\VRF=0.17$ which for the parameters chosen verifies that 
$p_\VRF > 2\Wb/(2\Wb+\Wp n)=1/6$, 
that $p_\VRF \geq \Wc/\Wb$ since $\Wc/\Wb\leq1/10$,
and that 
$p_\VRF \geq (e^{-n \xi^2/(6(1+\xi))} - \varphi)/(e^{-n \xi^2/(6(1+\xi))}-p_\CH^n)$, which is true for some $\varphi>0$ and $\xi>0$ as required by Theorem~\ref{thm:mixed}.
To implement the punishment decision in RG, we set the probability of punishment error to $\varepsilon=0.01$

For ROS, we set $p_\VRF=(\Wb+0.1)/(3\Wb)+0.01>(\Wb+\Wc)/(\Wp+2\Wb)$ as required by OS.
For the master payoffs, the aim in~\cite{ccpe} was to focus on the master cost making $\Vc=20$ but zeroing $\Mp$ and $\Mb$ to exclude the impact of whether the correct answer is obtained or not. Here, we consider such impact in all three mechanisms making $\Vc=\Mp=\Mb=n\Wb$, and we set $\Ac=\Wb$ assuming that the master cost of accepting an answer is just the payment that the worker receives (no overheads).

It is fair to notice that in ED the master checks the answers received by computing the task itself (\defn{audit}), which is usually more costly than just verifying a given solution (e.g., all NP problems that are not in P). 
Moreover, when the master audits, it obtains the correct result even if all workers cheat, which is not the case when verifying. Given that the master is penalized for obtaining a wrong answer, when verifying its utility is negatively affected with respect to auditing when no punishment is received by the master.
For these simulations, we maintain the same value for the master cost of verification or auditing, but in our model we zero the master punishment when all workers cheat (cf. Table~\ref{table:util} in the Appendix).

We assume that worker deviations occur in $0.5\%$ of the computation rounds, and that these deviations continue until ``fixed'' (if possible). That is, we evaluate the performance of all three mechanisms for $200$ rounds of computation introducing an identical initial perturbation in them. 
Specifically, we set $\lfloor n/2\rfloor$ workers to start with $p_\CH=0$ in ED and ROS, and $p_\CH$ as defined above for RG. For the remaining $\lceil n/2\rceil$ workers, we evaluate the range $p_\CH\in\{0.5,0.6,\dots,1\}$ in all three systems.
Notice that this number of deviators is minimal to have an impact on voting schemes.
ED will make the deviators converge to $p_\CH=0$ by design. ROS does not include previsions for deviators so they will not return to the desired behavior. For RG, we assume that, after being punished by peers in one round, workers return to the agreed equilibrium in the following round. 

Under such conditions, we compute the number of rounds when the master obtains the correct answer, the master and worker utilities aggregated over all these rounds, and we measure the convergence time for ED and the time to detect the deviation for RG. 
We discuss the results of our simulations in the following section.

\begin{figure*}[t]
       \centering
       {\label{fig:n9correctness}\resizebox{.47\textwidth}{!}{\input{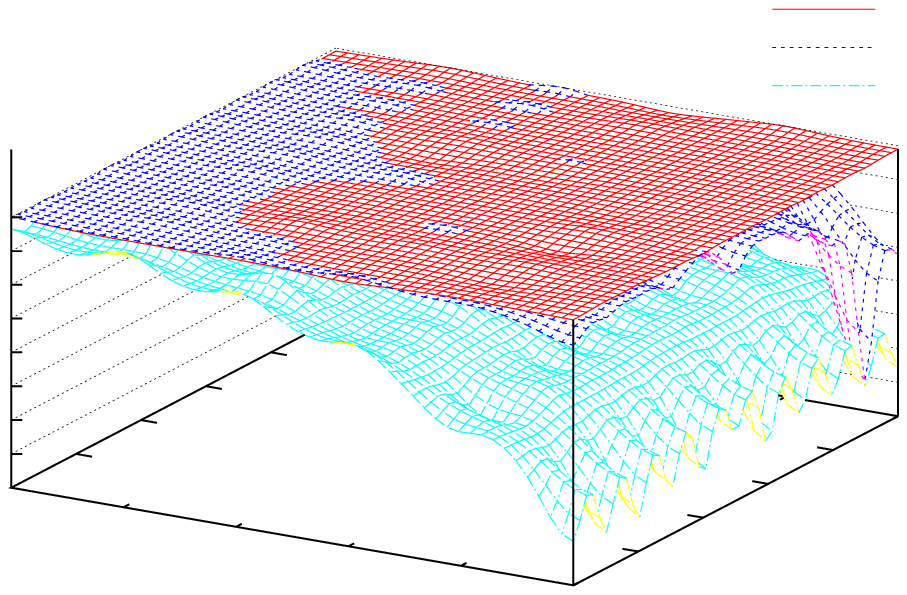}}}%
       {\label{fig:n9masterutil}\resizebox{.47\textwidth}{!}{\input{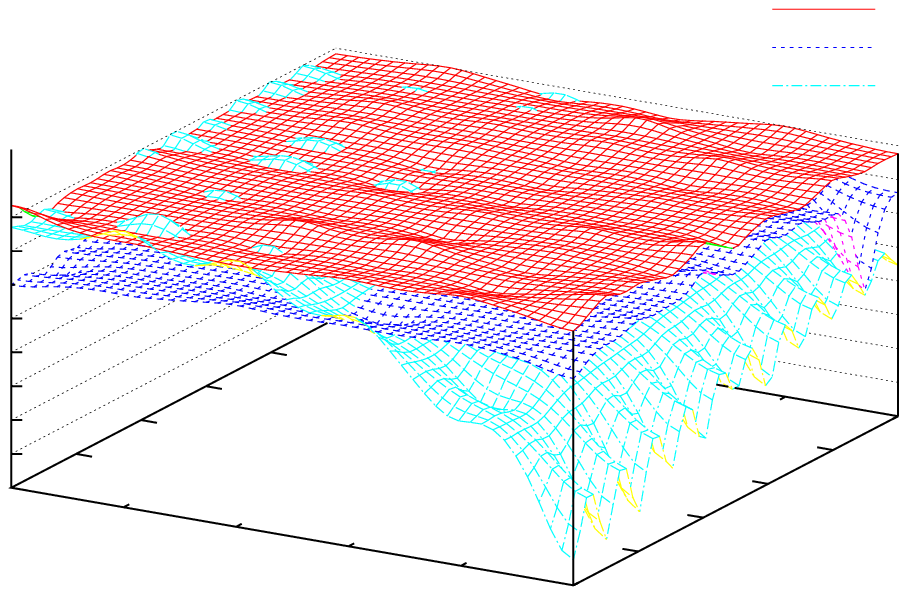}}}\\
       {\label{fig:n9workerutil}\resizebox{.47\textwidth}{!}{\input{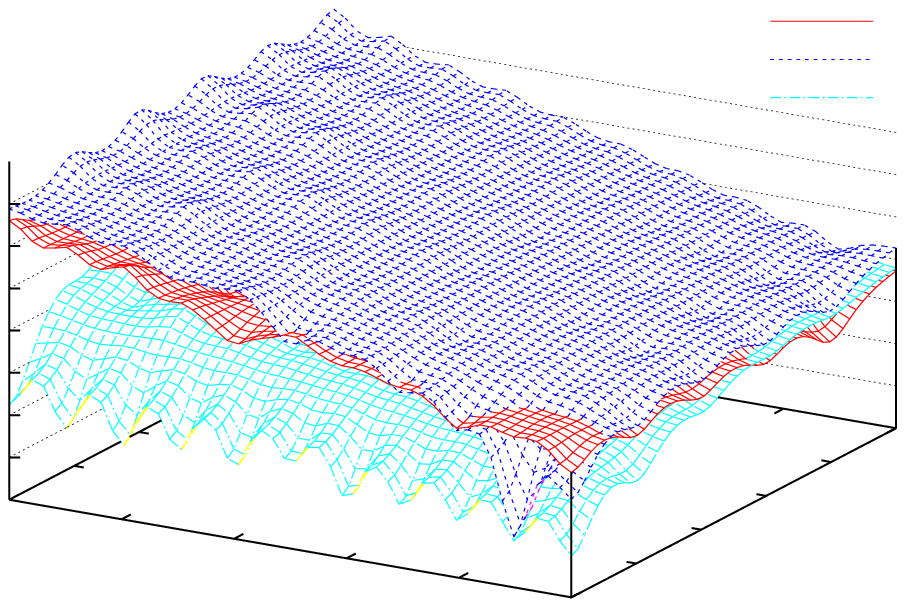}}}%
       {\label{fig:n9rounds}\resizebox{.47\textwidth}{!}{\input{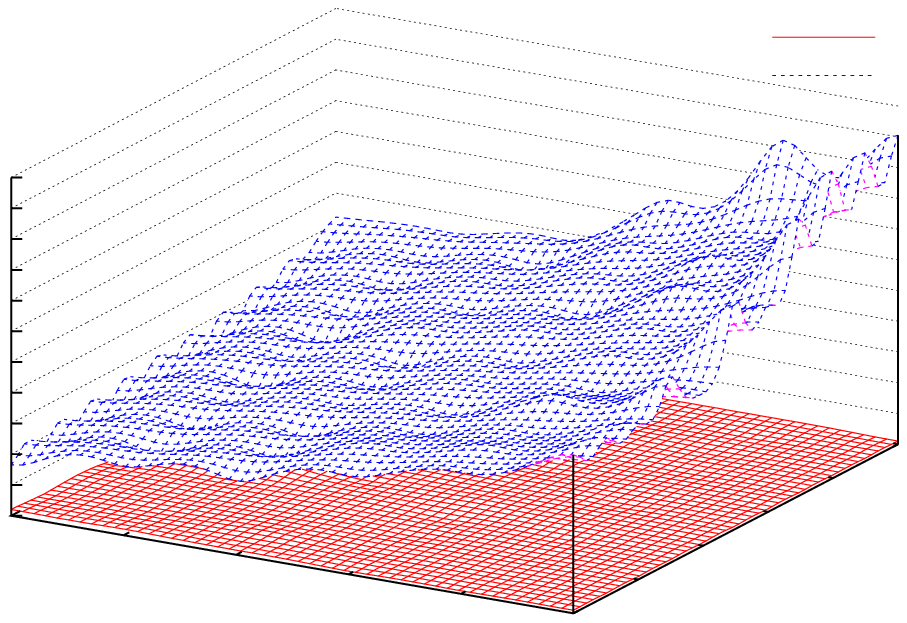}}}
       \caption[]{$n=9$}%
       \label{fig:n9}%
\end{figure*}

\subsection{Discussion}

The results of our simulations for $n=9$ are shown in Figure~\ref{fig:n9}. Similar results for $n=27$ and $81$ are shown in Figures~\ref{fig:n27} and~\ref{fig:n81}, left to the Appendix for brevity. The results shown correspond to one execution of the simulator. Multiple executions were carried out obtaining similar results.


It can be seen in Figure~\ref{fig:n9correctness} that the number of rounds when the master obtained the correct answer is similar for RG and ED, except when the deviator $p_\CH$ comes closer to $1$ where the performance of ED worsens. In general, both mechanisms achieve significantly better correctness than ROS. 

With respect to utilities, Figure~\ref{fig:n9masterutil} shows that RG is sensibly better than ED and ROS in master utility for most $(\Wb,p_\CH)$ combinations.
Yet, for follower-worker utility (Figure~\ref{fig:n9workerutil}), RG is almost as good as ED, which is slightly better because a follower worker in ED never cheats. Both, RG and ED, are significantly better than ROS where the deviator $p_\CH$ becomes bigger.

The intuition on why this performance is achieved by RG can be obtained from Figure~\ref{fig:n9rounds}, where it can be seen that, for these specific scenarios, our mechanism detects deviations very fast, in contrast with the slow convergence of ED. It should be noticed that fast detection of deviations does not necessarily imply correctness, given that in RG the equilibrium is some $p_\CH>0$. Yet, RG achieves correctness similar or better than ED, and much better than ROS, even though in the latter two the compliant (stable for ED) worker behavior is $p_\CH=0$.

\section{Conclusions}

Our simulations show that in presence of $\lceil n/2\rceil$ deviators, which is minimal to have an impact on voting schemes, even though the follower workers use $p_\CH>0$ (in contrast with the other mechanisms where $p_\CH=0$ for the followers), even assuming that the cost of verifying is the same as the cost of auditing, and even under the risk of unfair peer-punishment (i.e. workers may be punished even if they do not deviate, because the punishment decision is stochastic), our mechanism performs similarly or better than~\cite{ccpe}, and both significantly better than a repeated application of~\cite{plosone}.

These experimental results, together with our theoretical analysis validating the application of the repeated games framework, demonstrate the benefit and the promise of applying 
repeated games to the master-worker paradigm.
To the best of our knowledge, this is the first study of multi-round master-worker computing applying this framework.

A future extension of this work would be to enable the mechanism to also cope with malicious workers, that is, workers
that either intentionally or due to software or hardware errors, return an incorrect task result (recall the relative
discussion in the related section of the Introduction). Following~\cite{IEEETC14}, we could use statistical information on the distribution of the different worker types (malicious and rational). Then, the deviation threshold of our mechanism will need to be dependent on the expected number of malicious workers, so to keep motivating the rational workers to be truthful; we expect that the analysis will need to be significantly revised.  Another future extension would be, as in~\cite{plosone}, to consider the possibility of 
groups of workers colluding
in an attempt to 
increase their utility. For example, one worker could compute the task and inform the others of the correct result so that all return this result to
the master (and hence all would obtain the master's payment); or workers would return the same incorrect result in an attempt to cheat the master in accepting an incorrect task result. The challenge here is for the master to cope with these collusions, without knowing which specific workers are colluding.






\bibliographystyle{IEEEtran}
\bibliography{references,refs}

%
%
%

\appendix

\begin{figure*}[p]%
       \centering
       \subfloat[Number of correct rounds.][]{\label{fig:n27correctness}\resizebox{.47\textwidth}{!}{\input{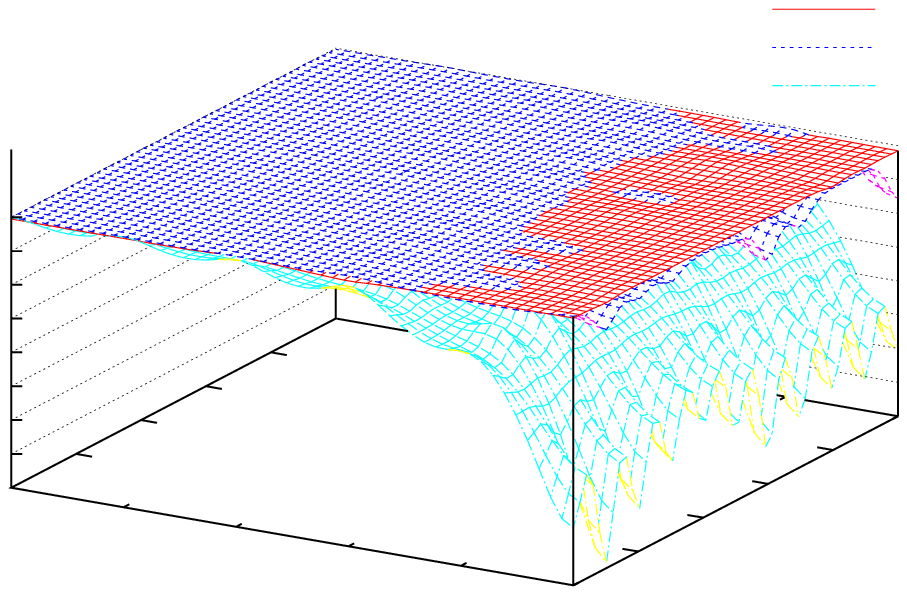}}}%
       \qquad
       \subfloat[Cumulative master utility.][]{\label{fig:n27masterutil}\resizebox{.47\textwidth}{!}{\input{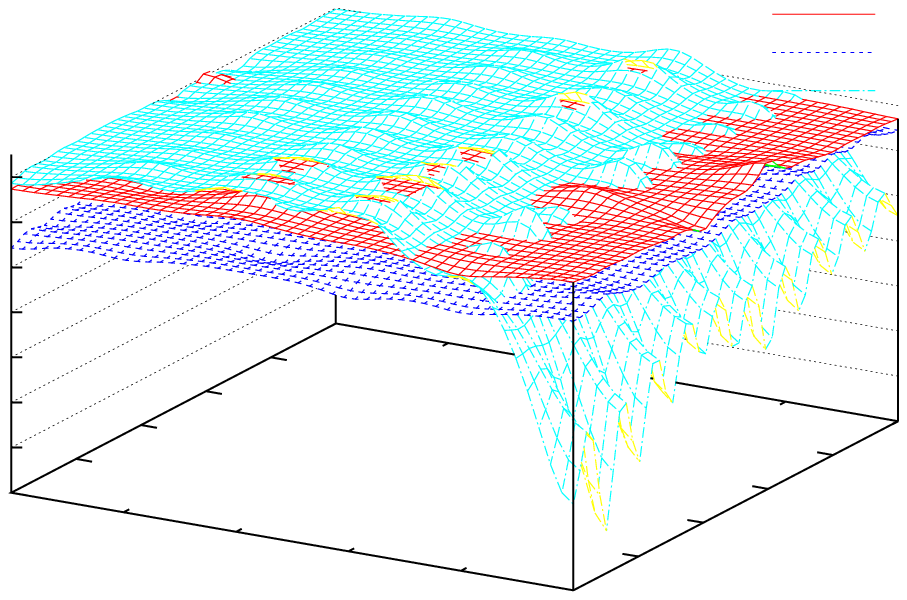}}}\\
       \subfloat[Cumulative follower worker utility.][]{\label{fig:n27workerutil}\resizebox{.47\textwidth}{!}{\input{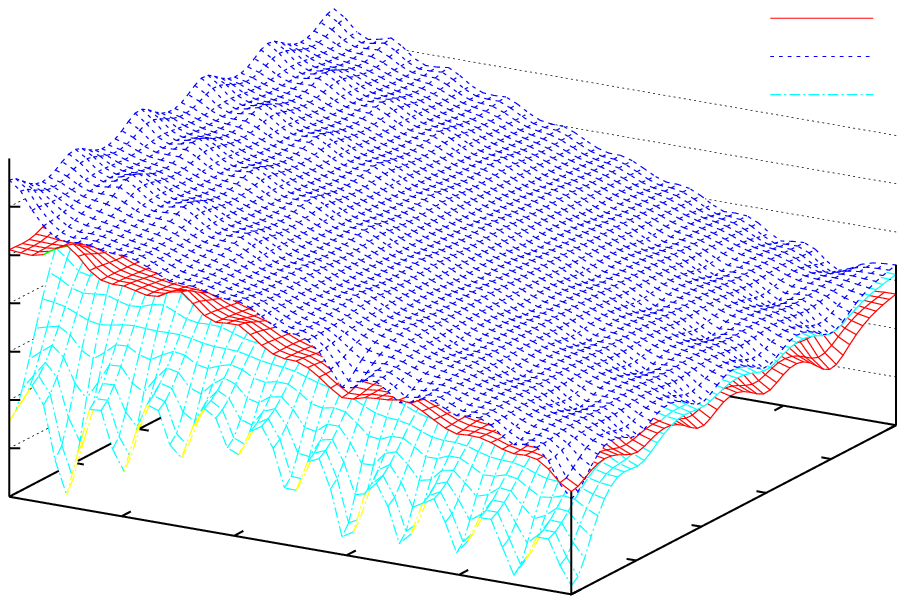}}}%
       \qquad
       \subfloat[Rounds to detection/convergence.][]{\label{fig:n27rounds}\resizebox{.47\textwidth}{!}{\input{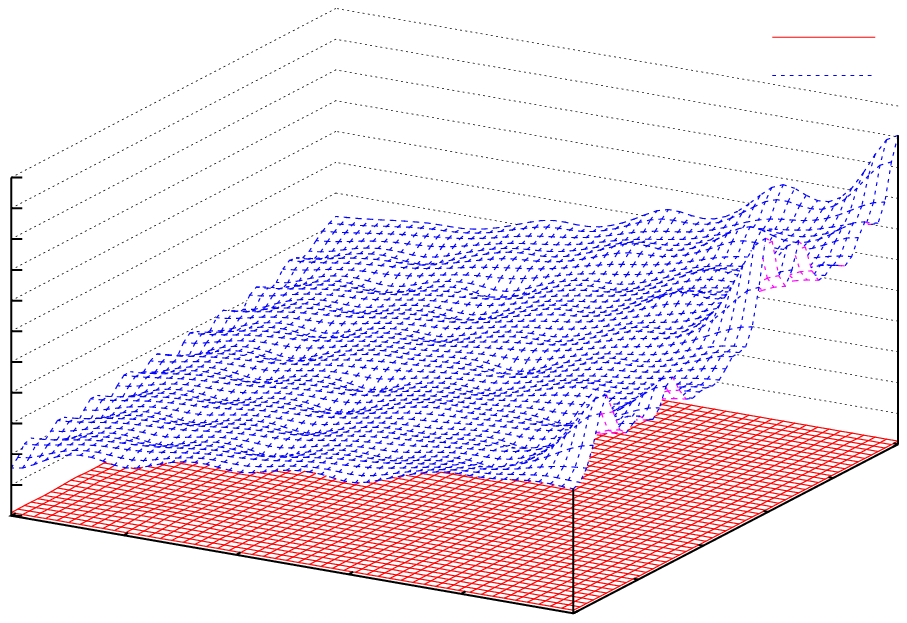}}}
       \caption[]{$n=27$}%
       \label{fig:n27}%
\end{figure*}

\begin{figure*}[p]%
       \centering
       \subfloat[Number of correct rounds.][]{\label{fig:n81correctness}\resizebox{.47\textwidth}{!}{\input{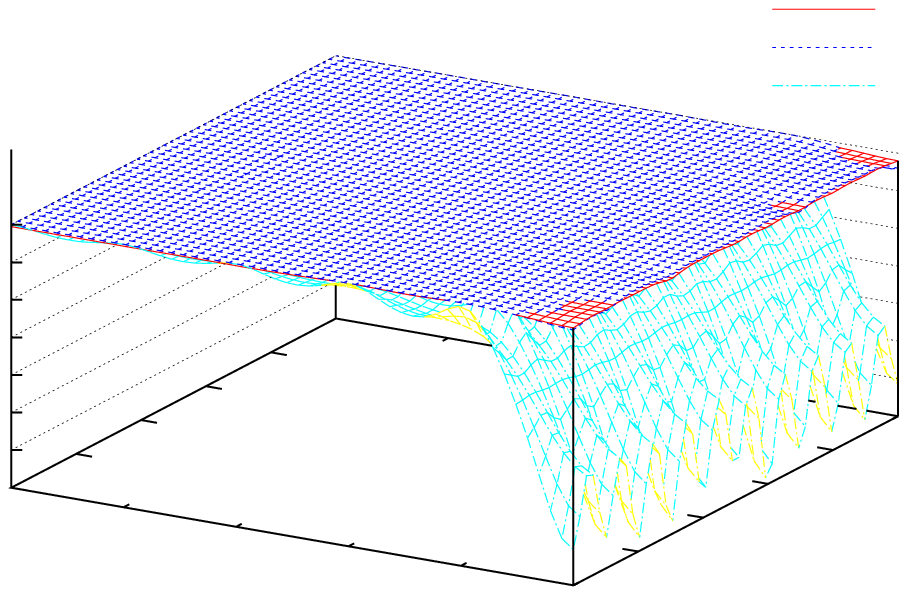}}}%
       \qquad
       \subfloat[Cumulative master utility.][]{\label{fig:n81masterutil}\resizebox{.47\textwidth}{!}{\input{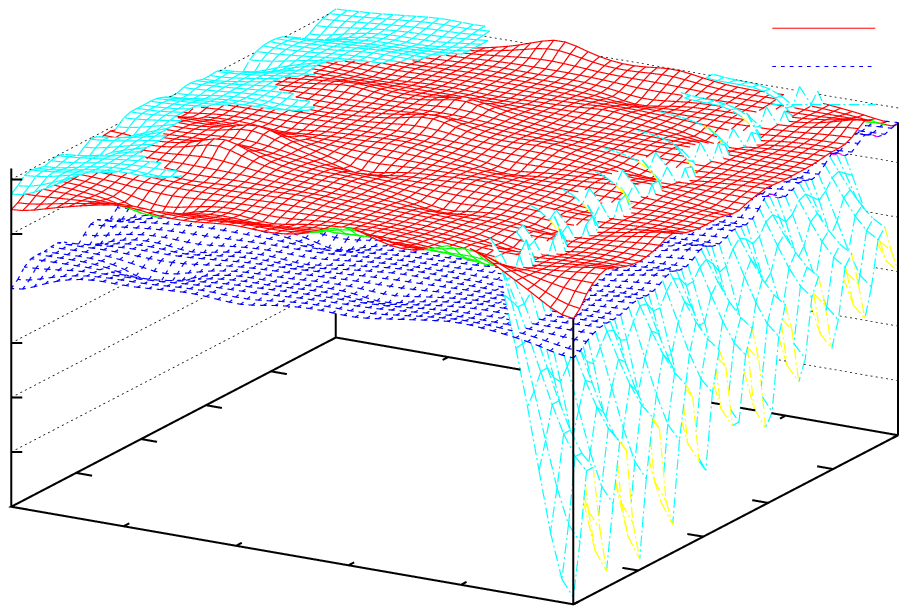}}}\\
       \subfloat[Cumulative follower worker utility.][]{\label{fig:n81workerutil}\resizebox{.47\textwidth}{!}{\input{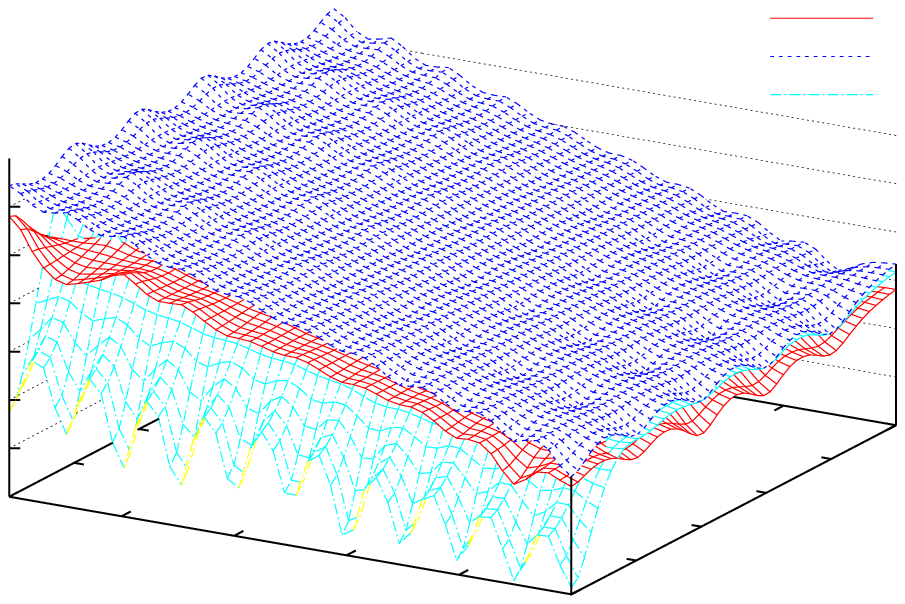}}}%
       \qquad
       \subfloat[Rounds to detection/convergence.][]{\label{fig:n81rounds}\resizebox{.47\textwidth}{!}{\input{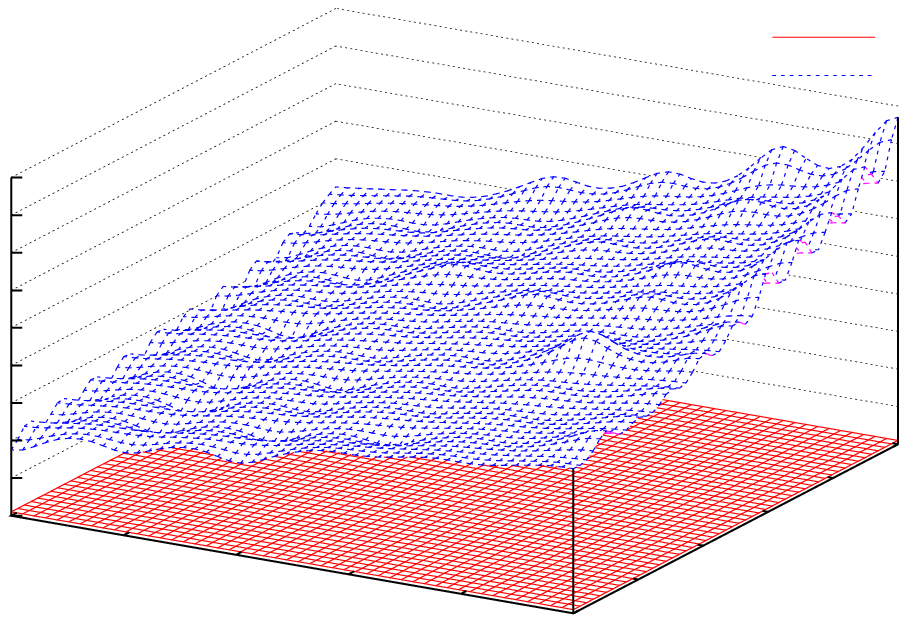}}}
       \caption[]{$n=81$}%
       \label{fig:n81}%
\end{figure*}

Table~\ref{table:param} summarizes the parameter values used in our simulations and Table~\ref{table:util} shows
the master and workers' utilities as derived under the specific parameter values. 
Figures~\ref{fig:n27} and~\ref{fig:n81} depict the simulation results for 27 and 81 workers, respectively.
As it can be observed, conclusions similar to those obtained for 9 workers can be derived here.

\begin{table*}[h]
\centering
\begin{tabularx}{\textwidth}{|c|Y|Y|Y|} 
\hline
& RG~[this paper] & ROS~\cite{plosone}& ED~\cite{ccpe}\\
\hline
$n$&$\{9,27,81\}$&$\{9,27,81\}$&$\{9,27,81\}$\\
\hline
$\Wb$&$\{1,1.1,1.2,\dots, 2\}$&$\{1,1.1,1.2,\dots, 2\}$&$\{1,1.1,1.2,\dots, 2\}$\\
\hline
$\Wp$&\begin{tabular}{c}$\Wb/(np_\CH)$ if $|F|<n$,\\ $0$ if $|F|=n$\end{tabular}&$\Wb$&$\Wb$\\
\hline
$\Wc$&$0.1$&$0.1$&$0.1$\\
\hline
$p_{\CH}$&\begin{tabular}{c}$\lfloor n/2\rfloor$:$0.1$,\\$\lceil n/2\rceil$:$\{0.5,0.6,\dots,1\}$\end{tabular}&\begin{tabular}{c}$\lfloor n/2\rfloor$:$0$,\\$\lceil n/2\rceil$:$\{0.5,0.6,\dots,1\}$\end{tabular}&\begin{tabular}{c}$\lfloor n/2\rfloor$:$0$,\\$\lceil n/2\rceil$:$\{0.5,0.6,\dots,1\}$\end{tabular}\\
\hline
$p_{\VRF}$&$0.17$&$(\Wb+0.1)/(3\Wb)+0.01$&\begin{tabular}{c}initially: $0.5$,\\ min: $0.01$\end{tabular}\\
\hline
$\Ac$&$\Wb$&$\Wb$&$\Wb$\\
\hline
$\Vc$&$n\Wb$&$n\Wb$&$n\Wb$\\
\hline
$\Mp$&$n\Wb$&$n\Wb$&$n\Wb$\\
\hline
$\Mb$&$n\Wb$&$n\Wb$&$n\Wb$\\
\hline
other&$\varepsilon=0.01$&--&\begin{tabular}{c}$\tau=0.5$,\\$a_w=0.1$,\\$\alpha_m=\alpha_w=0.01$\end{tabular}\\
\hline
\end{tabularx}
\caption{Simulations parameters.}
\label{table:param}
\end{table*}

\begin{table*}[h]
\centering
\begin{tabularx}{\textwidth}{|c|Y|Y|Y|} 
\hline
& RG~[this paper] & ROS~\cite{plosone}& ED~\cite{ccpe}\\
\hline
$U_i(\textrm{verified}, s_i=\NCH)$ & \multicolumn{3}{c|}{$\Wb-\Wc$}\\
\hline
$U_i(\textrm{verified}, s_i=\CH)$ & $- \Wp|F|$ &\multicolumn{2}{c|}{$- \Wp$}\\ 
\hline
$U_i(\textrm{not verified},s_i=\CH,|F|>n/2)$ & \multicolumn{3}{c|}{$\Wb$}\\
\hline
$U_i(\textrm{not verified}, s_i=\NCH,|F|>n/2)$ & \multicolumn{3}{c|}{$-\Wc$}\\
\hline
$U_i(\textrm{not verified}, s_i=\CH,|F|<n/2)$ & \multicolumn{3}{c|}{$0$}\\ 
\hline
$U_i(s_i=\NCH,|F|<n/2)$ & \multicolumn{3}{c|}{$\Wb-\Wc$}\\
\hline
$U_M(\textrm{verified},|F|<n)$ & \begin{tabular}{c}$\Mb-\Vc-$\\$(n-|F|)\Ac+|F|^2\Wp$\end{tabular} &\multicolumn{2}{c|}{$\Mb-\Vc-(n-|F|)\Ac+|F|\Wp$}\\
\hline
$U_M(\textrm{verified},|F|=n)$ & \multicolumn{2}{c|}{$-\Vc+n^2\Wp$}& $\Mb-\Vc+n\Wp$ \\
\hline
$U_M(\textrm{not verified},|F|>n/2)$ & \multicolumn{3}{c|}{$-\Mp-|F|\Ac$}\\
\hline
$U_M(\textrm{not verified},|F|<n/2)$ & \multicolumn{3}{c|}{$\Mb-(n-|F|)\Ac$}\\
\hline
\end{tabularx}
\caption{Master and worker utilities.}
\label{table:util}
\end{table*}

\end{document}